\documentclass[aps,prb,twocolumn,superscriptaddress]{revtex4}
\usepackage{eurosym}
\usepackage{amsfonts}
\usepackage{amssymb}
\usepackage{amsmath}
\usepackage{graphicx}
\usepackage{color}
\usepackage[hypertex,dvipdfm,colorlinks=true,urlcolor=blue,linkcolor=blue,citecolor=blue]{hyperref}
\begin{document}
\title{Manipulating charge-density-wave in monolayer $1T$-TiSe$_2$ by strain and charge doping: A first-principles investigation}
\author{M. J. Wei}
\affiliation{Key Laboratory of Materials Physics, Institute of Solid
State Physics, Chinese Academy of Sciences, Hefei 230031, China}
\affiliation{University of Science and Technology of China, Hefei 230026, China}

\author{W. J. Lu}
\email{wjlu@issp.ac.cn}
\affiliation{Key Laboratory of Materials
Physics, Institute of Solid State Physics, Chinese Academy of
Sciences, Hefei 230031, China}

\author{R. C. Xiao}
\affiliation{Key Laboratory of Materials Physics, Institute of Solid
State Physics, Chinese Academy of Sciences, Hefei 230031, China}
\affiliation{University of Science and Technology of China, Hefei 230026, China}

\author{H. Y. Lv}
\affiliation{Key Laboratory of Materials Physics, Institute of Solid
State Physics, Chinese Academy of Sciences, Hefei 230031, China}

\author{P. Tong}
\affiliation{Key Laboratory of Materials Physics, Institute of Solid
State Physics, Chinese Academy of Sciences, Hefei 230031, China}

\author{W. H. Song}
\email{whsong@issp.ac.cn}
\affiliation{Key Laboratory of Materials Physics, Institute of Solid
State Physics, Chinese Academy of Sciences, Hefei 230031, China}

\author{Y. P. Sun}
\affiliation{Key Laboratory of Materials Physics, Institute of Solid
State Physics, Chinese Academy of Sciences, Hefei 230031, China}
\affiliation{High Magnetic Field Laboratory, Chinese Academy of
Sciences, Hefei 230031, China}
\affiliation{Collaborative Innovation Center of Microstructures,
Nanjing University, Nanjing 210093, China }

\begin{abstract}
We investigate the effects of the in-plane biaxial strain and charge doping on the charge density wave (CDW) order of monolayer $1T$-TiSe$_2$  by using the first-principles calculations. Our results show that the tensile strain can significantly enhance the CDW order, while both compressive strain and charge doping (electrons and holes) suppress the CDW instability. The tensile strain may provide an effective method for obtaining higher CDW transition temperature on the basis of monolayer $1T$-TiSe$_2$. We also discuss the potential superconductivity in charge-doped monolayer $1T$-TiSe$_2$. Controllable electronic phase transition from CDW state to metallic state or even superconducting state can be realized in monolayer $1T$-TiSe$_2$, which makes $1T$-TiSe$_2$ possess a promising application in controllable switching electronic devices based on CDW.
\end{abstract}

\maketitle

\section{Introduction}
Layered transition metal dichalcogenides (TMDs) have received wide-spreading attentions, due to a variety of characteristic physical properties, such as superconducting, Mott insulating state and charge density wave (CDW) state.\cite{101, 100, 157} Among these rich physical properties, CDW is one of the fascinating collective phenomena in TMDs.\cite{10, 11} Below the CDW transition temperature, the lattice distorts simultaneously accompanied with the redistribution of charge and the abrupt change of electronic transport properties, which might open new potential applications in optoelectronic and quantum information processing devices.\cite{16, 13, 14, 18} For practice applications, the room temperature ultrathin CDW materials are desirable.

Previous experimental studies showed that reducing the thicknesses of the CDW materials can effectively tune the CDW transition temperature. For example, CDW transition temperature of $1T$-TaS$_2$ is shifted to lower temperature with the reduction of thickness and then disappears at a critical thickness.\cite{3503} Xi $et$ $al$. have reported that the CDW transition temperature of $2H$-NbSe$_2$ is enhanced from 33 K in the bulk to 145 K in the monolayer.\cite{140} Recently, Chen $et$ $al$. demonstrated that the CDW transition temperature of $1T$-TiSe$_2$ increases from 200 K in the bulk to 230 K in the monolayer.\cite{34} To the best of our knowledge, the CDW transition temperature of monolayer $1T$-TiSe$_2$ is comparatively closer to room temperature than that of other monolayer TMD materials. Hence, the monolayer $1T$-TiSe$_2$ provides an ideal platform for obtaining CDW order with higher transition temperature even above room temperature. This motivates us to search for methods to tune the CDW order of  monolayer $1T$-TiSe$_2$.

The ultrathin $1T$-TiSe$_2$ materials are often prepared on substrate, and the effects of the substrate-induced strain are inevitable, which offers opportunities for tuning CDW order in experiments.\cite{97, 3493} Meanwhile, charge doping has also an important impact on the CDW transition.\cite{67, 15} Hence, in this work, we focus on the effects of the strain and charge doping on the CDW order in monolayer $1T$-TiSe$_2$ by using the first-principles calculations. Our results show that the tensile strain can enhance the CDW order, while the compressive strain and charge doping suppress it. The CDW gap increases with the increase of tensile strain, while the compressive strain reduces CDW gap and it undergoes a semiconductor-metal transition at 6\% compressive strain. Furthermore, we discuss that the superconductivity with superconducting transition temperature $T_C$ of 7.3-0.3 K can be introduced by electron/hole doping.

\section{COMPUTATIONAL DETAILS}
The first-principles calculations were carried out by using the QUANTUM ESPRESSO package\cite{156} with ultra-soft pseudopotentials.\cite{43} The generalized gradient approximation according to Perdew-Burke-Ernzerhof (PBE) \cite{44} functional was used. The valence electrons were simulated by plane wave method. The energy cutoff of 80 Ry (800 Ry) was chosen for the wave functions (charge density) basis. In order to simulate the monolayer, at least 18 $\mathrm\AA$ of vacuum layer was introduced. The Brillouin zone (BZ) was sampled on a $16\times16\times1$ mesh of \emph{\textbf{k}}-points. The total energy and electron charge density were calculated by the Fermi-Dirac smearing method. If there are no special notes, a smearing parameter of $\sigma$ was set to $\sigma = 0.005$ Ry. Using density functional perturbation theory (DFPT),\cite{3462} phonon dispersion curves were calculated with $8\times8\times1$ mesh of \emph{\textbf{q}}-points for the monolayer sample. Denser $32\times32\times1$ mesh of \emph{\textbf{k}}-points were used in the electron-phonon coupling calculations. For a better agreement with the experimental results, LDA$+U$ method with the value $U$=3.9 eV\cite{45} and spin-orbit interactions were considered in the electronic structure calculations. Since the spin-orbit coupling effect is less important in describing the vibrational properties,\cite{3509, 3508} the calculation of phonon dispersion was carried out neglecting this effect. The biaxial strain was simulated by changing the lattice constant $a_0$, and the strength of strain was defined as $\varepsilon = (a-a_0)/a_0\times100\%$, for which the positive (negative) value represents the tensile (compressive) strain. For the investigation of the charge doping effects, the electron doping was simulated by increasing electrons into the system, while the hole doping by removing electrons from the system, together with compensating uniform positive background for electron doping and negative background for hole doping. For each doping concentration, atomic positions were relaxed with fixing the lattice parameters of the undoped monolayer $1T$-TiSe$_2$.

\section{Results and discussion}

Bulk $1T$-TiSe$_2$ exhibits a layered structure with the space group $P$\=3m1, where adjacent layers are held together by van der Waals forces and each Ti atoms are surrounded by the nearest six Se atoms, constituting an octahedron, as illustrated in Fig. 1(a). The monolayer $1T$-TiSe$_2$ can be obtained by exfoliating the bulk $1T$-TiSe$_2$ or grown by molecular beam epitaxy.\cite{15, 34} Below ~230 K, the monolayer $1T$-TiSe$_2$ undergoes a $2\times2\times1$ commensurate CDW transition,\cite{34} so primitive cell lattice doubles as displayed in Fig. 1(b), and the corresponding BZ shrinks half of that of the undistorted monolayer $1T$-TiSe$_2$.

\begin{figure}
\begin{flushleft}
\includegraphics[width=0.95\columnwidth]{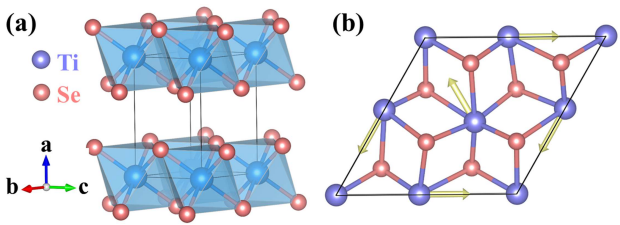}
\caption{Crystal structures of bulk $1T$-TiSe$_2$ for (a) normal phase at room temperature, (b) $2\times2\times1$ CDW phase at low temperature. Yellow arrows denote the displacements directions of Ti atoms from the high symmetry positions of the undistorted 1$T$ structure. Ti and Se atoms are denoted by blue and red balls, respectively.}
\end{flushleft}
\end{figure}

When the bulk is thinned to the monolayer, the strain can be an effective method to manipulate the properties of materials. Hence, we firstly investigate the evolution of the CDW order in monolayer $1T$-TiSe$_2$ under the in-plane biaxial compressive and tensile strains. In order to comprehend the stability of CDW order from an energy perspective, the CDW formation energy $\Delta E$ is defined as:
\begin{equation}
\Delta E=E_{CDW}-E_{1T},
\end{equation}
where $E_{CDW}$ and $E_{1T}$ are the total energies of the relaxed CDW structure and the undistorted $1T$ structure. Meanwhile, the CDW phase transitions can be ascribed to the results of spontaneous symmetry breaking, which involves atomic displacements, as shown in Fig. 1(b).\cite{34} Hence, the average displacements of Ti atoms, D$_{Ti}$, have been introduced to describe the strain and charge doping dependence of CDW order from a structural perspective.\cite{50, 49} Figure 2 shows the CDW formation energy and the average displacements of Ti atoms of the monolayer $1T$-TiSe$_2$ in the CDW phase as a function of biaxial strain. When the tensile strain was applied to monolayer $1T$-TiSe$_2$, the absolute value of the formation energy gradually increases, indicating that the CDW phase becomes more stable. On the contrary, the compressive strain gradually suppresses the CDW instability. From a structural perspective, the distortion amplitude of Ti atoms is remarkably enhanced by the tensile strain, while the compressive strain weakens it. These results indicate that the biaxial compressive and tensile strain can remarkably tune the CDW order.
\begin{figure}
\begin{flushleft}
\centering\includegraphics[width=0.95\columnwidth]{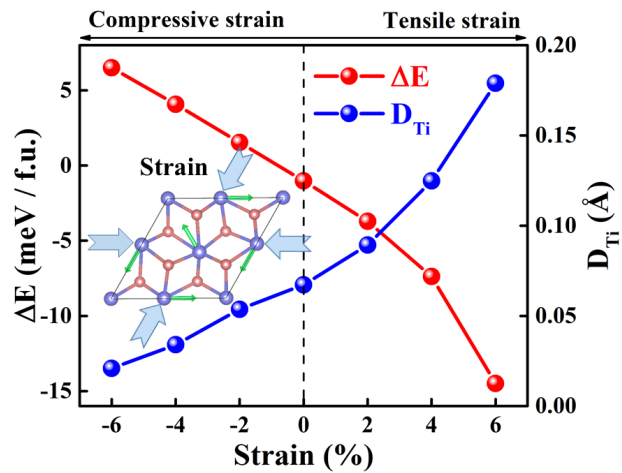}
\caption{CDW formation energy $\Delta E$  per formula unit (left) and average displacements of Ti atoms from the high symmetry positions of the undistorted $1T$ structure (right) as a function of biaxial strain. The inset shows the biaxial compressive strain on the monolayer $1T$-TiSe$_2$ in CDW phase.}
\end{flushleft}
\end{figure}

\begin{figure*}
\includegraphics[width=0.8\textwidth]{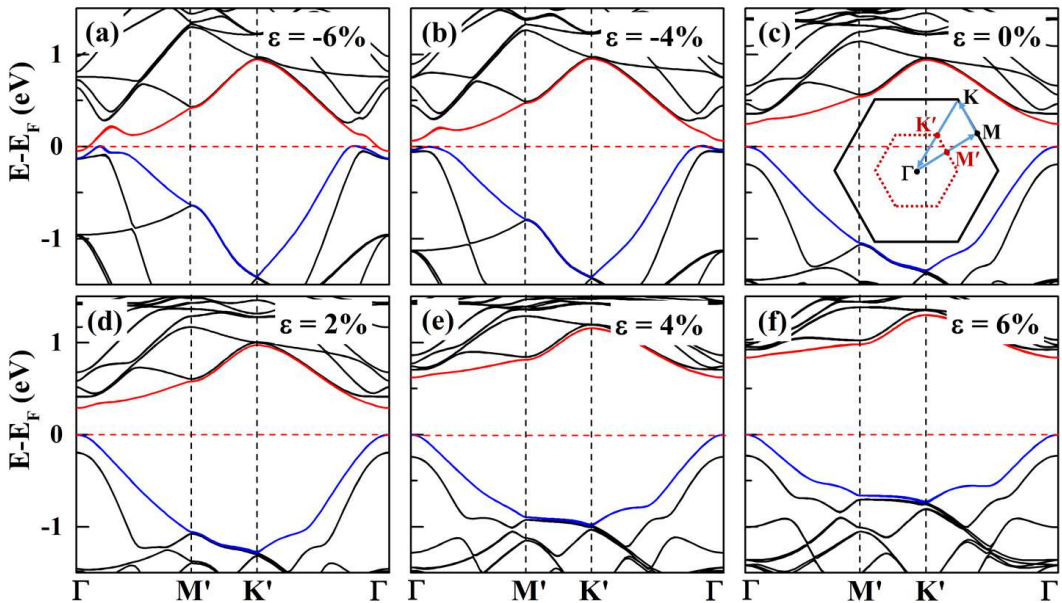}
\caption{Evolution of the electronic band structures of the monolayer $1T$-TiSe$_2$ in CDW phase under biaxial strain. Blue (red) solid lines in each panel indicates the highest valence band (the lowest conduction band). The top of valence band is set to zero. The inset in (c) shows the BZs of the monolayer $1T$-TiSe$_2$: black solid line represents the BZ of normal phase and red dotted line represents the folded BZ due to CDW phase transition.}
\end{figure*}

Figure 3 shows the electronic band structures of the monolayer $1T$-TiSe$_2$ in CDW phase as a function of biaxial strain. For comparison, the electronic band structures without applying strain are also shown (see Fig. 3(c)). We find that the band gap has different responses to the tensile and compressive strains. The tensile strain significantly increases the band gap, while the compressive strain reduces it. A continuously varying band gap can be achieved by strain. Therefore, such behavior leads to the transition from semiconductor to metal at 6\% compressive strain, as shown in Fig. 3(a).

The phonon instability of high-symmetry structure can be described as a result of the CDW distortion: closing to but higher than the CDW transition temperature, the phonon frequencies in the vicinity of the CDW vector \emph{\textbf{q}}$_{CDW}$ softens but does not go to zero, while the phonon frequencies become imaginary below the CDW transition temperature.\cite{113} The occurrence of imaginary phonon frequencies implies that the structure is unstable. A structure is considered to be stable when the calculated frequencies of all phonon modes are positive. Hence, the phonon calculations offer an effective method to simulate the CDW instability. Figure 4(c) shows the calculated phonon dispersion curves of monolayer $1T$-TiSe$_2$. We find that the imaginary/soften acoustic mode at $M$ point associated with the $2\times2\times1$ CDW instability of monolayer $1T$-TiSe$_2$ becomes more dynamically unstable than that of bulk (data not shown here), indicating that the CDW order of monolayer $1T$-TiSe$_2$ is more robust than that of the bulk. This is qualitatively consistent with the experimental observations that reducing thickness increases CDW transition temperature.\cite{18, 34, 3461}

\begin{figure*}
\includegraphics[width=0.8\textwidth]{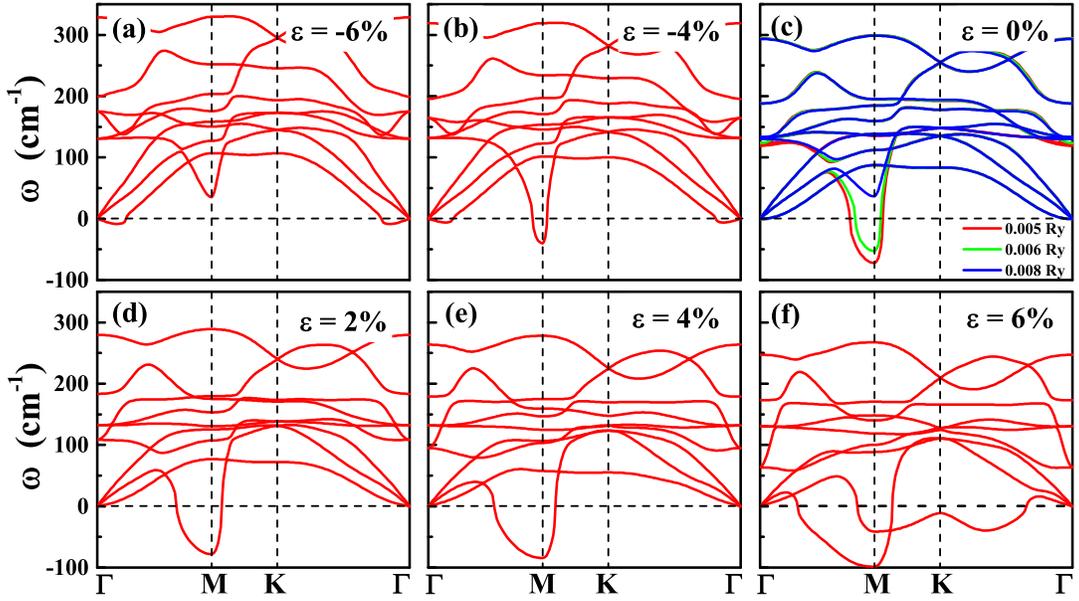}
\caption{Evolution of the phonon dispersion curves of monolayer $1T$-TiSe$_2$ in the normal phase under the biaxial strain. (c) The phonon dispersion curves as a function of $\sigma$.}
\end{figure*}

In order to further understand the evolution of the CDW order under the biaxial strain, we calculated the phonon dispersion curves under different biaxial strains. Figures 4(a)-(c) show that the compressive strain can enhance the phonon frequencies and reduce the area of instability at the CDW vector \emph{\textbf{q}}$_M$. When the compressive strain of 6\% is applied, the previous imaginary frequencies at $M$ point become positive (see Fig. 4(a)), indicating that the normal phase becomes stable against CDW transition. Figures 4(c)-(f) show that the tensile strain can reduce the phonon frequencies and largely expand the area of instability at the CDW vector \emph{\textbf{q}}$_M$. When the tensile strain of 6\% is applied, additional imaginary acoustic branches are introduced (see Fig. 4(f)), indicating that the larger tensile strain results in the structure phase transition in monolayer $1T$-TiSe$_2$. We can conclude that the tensile strain can enhance the CDW order, while the compressive strain suppresses it. It is expected that the tensile strain could further increase the CDW transition temperature of monolayer $1T$-TiSe$_2$. In addition, we note that there are small imaginary frequencies closing to $\Gamma$  point in Figs. 4(a) and 4(b), which is consistent with the instability against the long-wavelength transverse waves.\cite{53, 55} Such small instability is suggested to be fixed by defects, such as ripples or grain boundaries.\cite{53, 55, 54}

In order to qualitatively simulate the influence of temperature on CDW order, the temperature dependence of phonon dispersion curves were calculated by changing the smearing factor $\sigma$. Such $\sigma$ takes on a physical meaning to directly reflect the electronic temperatures of the system.\cite{56} The disappearance of the imaginary phonon frequencies at $M$ point in phonon dispersion curves means that the CDW instability is eliminated by increasing $\sigma$. Figure 4(c) shows the phonon dispersion curves of the unstrained monolayer $1T$-TiSe$_2$ with different $\sigma$. The imaginary/soften acoustic branches shows significantly dependence on the electronic temperature. As the electronic temperature increases, the area of the instability reduces and disappears at $\sigma$ = 0.008 Ry. According to the method proposed by Duong $et$ $al$.,\cite{56} the CDW transition temperatures $T_{CDW}$ can be obtained by fitting the smearing temperature-phonon frequencies at $M$ point based on the equation: \cite{56}
\begin{equation}
\omega_M=\omega_0^* (T-T_{CDW})^\delta.
\end{equation}
Figures 5(a) and 5(b) show the fitted smearing temperature dependence of phonon frequencies at $M$ point and the obtained $T_{CDW}$ under tensile strain, respectively. As the tensile strain increases, the $T_{CDW}$  becomes higher, further demonstrating that the tensile strain can enhance the CDW order. We can conclude that the control of  strain allows us to manipulate the CDW transition, which may open a promising application in future to construct controllable switching electronic devices based on CDW. Note that the real temperature of the crystal should include not only the temperature of electron but also of lattice, hence the calculated electron temperature cannot directly represent the real physical temperature and needs to be further amended. Despite the fact that the obtained $T_{CDW}$ is not accurate enough numerically, it can correctly describe the evolution of the CDW transition under the strain.
\begin{figure}
\begin{flushleft}
\includegraphics[width=0.95\columnwidth]{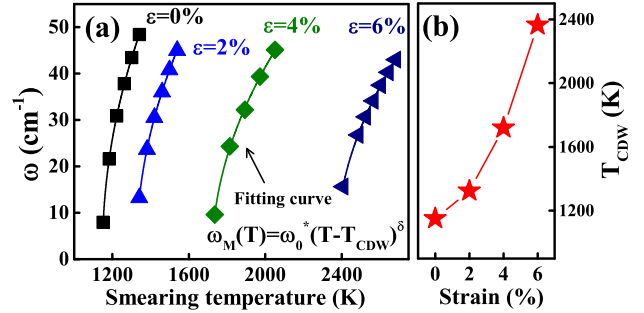}
\caption{(a) Phonon frequencies of soften acoustic mode at $M$ point as a function of the electronic temperature under tensile strain. The solid lines are fitted by Eq. (2). (b) Fitted CDW transition temperatures $T_{CDW}$ under tensile strain.}
\end{flushleft}
\end{figure}

Charge carrier doping can be easily introduced in experiments, such as gate-controlled Li ion intercalation, in-plane electric-field, and photoexcitation, which has been successfully applied to manipulate the CDW transition.\cite{3490, 3491, 3492} Here, we also investigated the effect of the charge doping on the CDW order in monolayer $1T$-TiSe$_2$. Figures 6(a) and 6(b) show the calculated phonon dispersion curves of monolayer $1T$-TiSe$_2$ with different doping concentrations. They show that both electron and hole doping can suppress the CDW instability. In order to more clearly see the evolution of the suppression of the CDW transition by charge doping, a series of phonon dispersion curves of monolayer $1T$-TiSe$_2$ and the average distortion of Ti atoms in CDW-phase TiSe$_2$ under different doping concentrations were calculated. The phonon frequencies at $M$ and the average distortion of Ti atoms were summarized in Fig. 6(c). One can see that the average distortion of Ti atoms gradually becomes smaller and eventually tends to zero with the increase of doping concentrations. When the doping concentration is above $n$=0.165 electrons/f.u. or $n$=0.055 holes/f.u., the previous imaginary frequencies at $M$ become positive, which indicates the doped monolayer $1T$-TiSe$_2$  with normal phase becomes completely stable. We can conclude that both electron and hole doping are harmful to stabilize the CDW order in the system. Hence, if one wants to obtain high CDW transition temperature in monolayer $1T$-TiSe$_2$, charge doping needs to be prohibited.

\begin{figure}
\begin{flushleft}
\includegraphics[width=0.95\columnwidth]{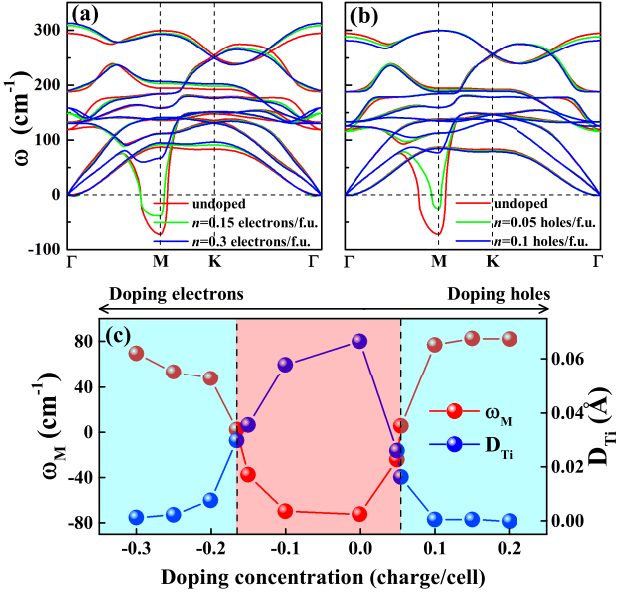}
\caption{Phonon dispersion curves for (a) electron doped and (b) hole doped monolayer $1T$-TiSe$_2$ under some typical doping concentrations. For comparison, undoped monolayer $1T$-TiSe$_2$ is also shown, which is depicted as red solid lines. (c) The phonon frequencies at $M$  point (left) and the average displacements of Ti atoms from the high symmetry positions of the undistorted $1T$ structure (right) as a function of doping concentration. The region decorated with red color stands for the CDW phase.}
\end{flushleft}
\end{figure}

Previously, the experiment reported that the CDW in $1T$-Cu$_x$TiSe$_2$ is continuously suppressed with the increase of Cu intercalation concentration and a superconducting state appears with a maximum superconducting transition temperature $T_C$ of 4.1 K at $x$=0.08.\cite{3502} The Cu intercalation is equivalent to introducing the electron into the system. Here, it is natural to ask whether pure charge doping can induce the potential superconductivity in monolayer $1T$-TiSe$_2$, when the CDW is suppressed. In order to evaluate the superconductivity, we estimate the $T_C$ by using the Allen-Dynes-modified McMillan formula:\cite{3486, 3485}
\begin{equation}
T_c=\frac{\omega_{log}}{1.2}\exp\left(-\frac{1.04(1+\lambda)}{\lambda-\mu^*-0.62\lambda\mu^*}\right) ,
\end{equation}
where the total electron-phonon coupling constant is
\begin{equation}
\lambda=\sum_{\mathbf{q} v}\lambda_{\mathbf{q} v}
=2\int\frac{\alpha^2F(\omega)}{\omega}\mathrm{d}\omega.
\end{equation}
The Coulomb pseudopotential $\mu^*$ is generally assumed to 0.1.\cite{94, 3488, 3487} The logarithmic average frequency of $\omega_{log}$  is defined as:
\begin{equation}
\omega_{log}=\exp\left(\frac{2}{\lambda}\int\frac{d\omega}{\omega}\alpha^{2}F(\omega)\log\omega\right),
\end{equation}
with the Eliashberg spectral function
\begin{equation}
  \alpha^2F(\omega)=\frac{1}{2\pi N(E_F)}\sum_{\mathbf{q}v}
  \delta(\omega-\omega_{\mathbf{q}v})\frac{\gamma_{\mathbf{q}v}}{\hbar \omega_{\mathbf{q}v}},
\end{equation}
where $N(E_F)$ is the density of states at $E_F$, and ${\gamma_{\mathbf{q}v}}$  is the phonon linewidth.

\begin{figure}
\begin{flushleft}
\includegraphics[width=0.95\columnwidth]{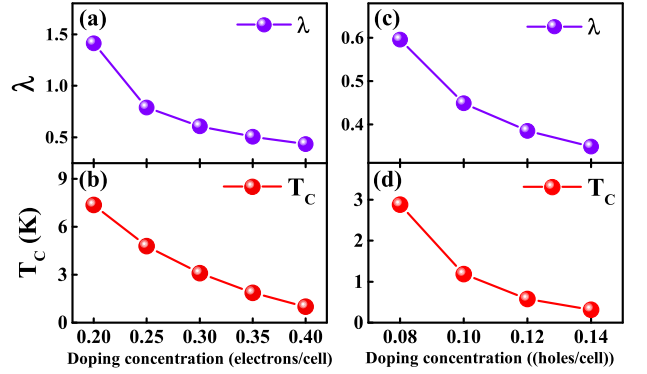}
\caption{Calculated  $\lambda$ and $T_C$ of monolayer $1T$-TiSe$_2$ under the (a and b) electron doping and (c and d) hole doping.}
\end{flushleft}
\end{figure}
Figures 7(b) and 7(d) show that the doped monolayer $1T$-TiSe$_2$ could be superconducting and the $T_C$ decreases monotonously
from 7.3 K (2.8 K) at 0.2 electrons/f.u. (0.08 holes/f.u.) to 0.9 K (0.3 K) at 0.4 electrons/f.u. (0.14 holes/f.u.). We try to understand the evolution: as the CDW instability is just suppressed and the system enters into the normal phase, the original imaginary frequency $\omega$ at $M$ point becomes a small positive value. The small positive $\omega$ in Eq. (4) can lead to a large $\lambda$, resulting in the high $T_C$. However, according to Fig. 6(c), both electron doping and hole doping make the phonon spectra harden, which leads to the reduction of $\lambda$ and is not beneficial to the superconductivity (Figs. 7(a) and (c)). As a result, the $T_C$ decreases monotonously with the increase of the charge doping concentration.
\section{CONCLUSIONS}
In conclusion, we studied the effects of the biaxial strain and charge doping on the CDW order of monolayer $1T$-TiSe$_2$ by the first-principles calculations. We found that the tensile strain can effectively enhance the CDW order. The results indicate that the CDW transition temperature of monolayer $1T$-TiSe$_2$ grown on substrates with large lattice constants can be enhanced. The compressive strain and the charge doping can suppress the CDW order. At the 6\% compressive strain or at the doping concentrations of 0.165 electrons/f.u. and 0.055 holes/f.u., the CDW instability is completely eliminated. When the CDW instability is suppressed by charge doping, the superconductivity with $T_C$ of 7.3-0.3 K can be introduced by electron/hole doping. Controllable electronic phase transition from CDW state to metallic state or even superconducting state can be realized in monolayer $1T$-TiSe$_2$, which makes $1T$-TiSe$_2$ possess a promising application in controllable switching electronic devices based on CDW.

\section*{ACKNOWLEDGEMENTS}
This work was supported by the National Key Research and Development Program of China under Contract No. 2016YFA0300404, the National Nature Science Foundation of China under Contract No. 11674326, 11404340, 11274311, and U1232139, Key Research Program of Frontier Sciences of CAS (QYZDB-SSW-SLH015), Hefei Science Center of CAS (2016HSC-IU011) and Youth Innovation Promotion Association of CAS (2012310). The calculations were partially performed at the Center for Computational Science, CASHIPS.

\end {document}